\documentclass[superscriptaddress]{revtex4-2}

\usepackage{graphicx}
\usepackage{color}
\usepackage[english]{babel}
\usepackage{rotating}
\usepackage{braket}
\usepackage{multirow}
\usepackage{makecell}
\usepackage{longtable}
\usepackage{amsmath}

\usepackage{hyperref}

\graphicspath{{./img/}}

\begin{document}

\title{Quantum Sensing of Magnetic Fields with Molecular Spins}

\author{Claudio Bonizzoni}
\email[mail to:]{ claudio.bonizzoni@unimore.it}
\affiliation{Dipartimento di Scienze Fisiche, Informatiche e Matematiche Universit\`a di Modena e Reggio Emilia, via G. Campi 213/A, 41125, Modena, Italy}
\affiliation{CNR Istituto Nanoscienze, Centro S3, via G. Campi 213/A, 41125, Modena, Italy.}
\author{Alberto Ghirri}
\affiliation{CNR Istituto Nanoscienze, Centro S3, via G. Campi 213/A, 41125, Modena, Italy.}
\author{Fabio Santanni}
\affiliation{Dipartimento di Chimica Ugo Schiff, via della Lastruccia 3, 50019 Sesto Fiorentino (FI), Italy.}
\author{Marco Affronte}
\affiliation{Dipartimento di Scienze Fisiche, Informatiche e Matematiche Universit\`a di Modena e Reggio Emilia, via G. Campi 213/A, 41125, Modena, Italy}
\affiliation{CNR Istituto Nanoscienze, Centro S3, via G. Campi 213/A, 41125, Modena, Italy.}

\date{\today}

\begin{abstract}
Spins are prototypical systems with the potential to probe magnetic fields down to the atomic scale limit. Exploiting their quantum nature through appropriate sensing protocols allows to enlarge their applicability to fields not always accessible by classical sensors. 
Here we first show that quantum sensing protocols for AC magnetic fields can be implemented on molecular spin ensembles embedded into hybrid quantum circuits. We then show that, using only echo detection at microwave frequency and no optical readout, Dynamical Decoupling protocols synchronized with the AC magnetic fields can enhance the sensitivity up to $S \approx 10^{-10}-10^{-9}\,\text{T}/\sqrt{\text{Hz}}$ with a low (4-5) number of applied pulses. These results paves the way for the development of strategies to exploit molecular spins as quantum sensors. 
\end{abstract}

\maketitle

\section{Introduction}
\label{sec.intro}

Quantum sensing, \textit{i.e.} the exploitation of genuine quantum properties such as quantum coherence or entanglement to probe tiny physical quantities, can overcome fundamental limitations or even outperform classical sensing \cite{degenREVMODPHYS2017}. Its advantages have been proved by measuring a large variety of physical quantities, such as electric or magnetic fields, temperature, pressure, rotation angle or speed, time (frequency) and acceleration \cite{degenREVMODPHYS2017}. This has brought to applications in photonics \cite{pirandolaNATPHOT2018}, nuclear \cite{allertCHEMCOMM2022,holzgrafePRAPPL2020} and electron spin resonance \cite{wolfowiczNATREVMAT2021,schirhaglANNREVPHYSCHEM2014,rondinREPPROGRPHYS2014,abeJAP2018,segawaPROGRNUCLMAGNRES2023} among others. 
Spins are natural candidates to perform quantum sensing of magnetic fields. Recently, Nitrogen-Vacancy (NV) centers in diamonds, both as ensembles (bulk collections) as well as single (isolated) spins have shown remarkable performances as quantum sensors \cite{schirhaglANNREVPHYSCHEM2014}. Sensitivities on the order of $\mu \text{T} /\sqrt{\text{Hz}}$ have been reported for single NV centers used as tips in scanning probe microscopy \cite{huxterNATCOMM2022,pelliccioneNATNANO2016,schirhaglANNREVPHYSCHEM2014}, while typical values from few units and up to tens of $n \text{T} /\sqrt{\text{Hz}}$ are reported for single spins using Optically Detected Magnetic Resonance (ODMR) protocols based on Dynamical Decoupling MW sequences \cite{wrachtrupJMR2016,taylorNATPHYS2011,balasubramanianNATMAT2009}. 
Using ensembles has the advantage to give the parallel average of the response of $N$ (identical) copies of single spins \cite{zhouPRX2020} and larger fluorescence signals to be measured in ODMR \cite{taylorNATPHYS2011,zhouPRX2020}. However, increasing the spin concentration reduces the memory time resulting in an upper bound for sensitivity, which can be overcame only with very specific and ad-hoc microwave decoupling schemes \cite{zhouPRX2020}. Typical sensitivity values of the order of units or tens of $n \text{T} /\sqrt{\text{Hz}}$ have been achieved on bulk sensors by means of ODMR combined with long (tens to hundred/s of pulses) Dynamical Decoupling sequences such as Carr-Purcell-Meiboom-Gill or XY8 \cite{phamPRB2012,rondinREPPROGRPHYS2014,taylorNATPHYS2011}, concatenated Dynamical Decoupling \cite{farfurnikJOURNOFOPT2018}, or advanced protocols compensating sample-specific inhomogeneities and disorder \cite{zhouPRX2020,balasubramanianNANOLETT2019}. Record sensitivity of $p \text{T} /\sqrt{\text{Hz}}$ with a sensor of $10^{13}$ spins and a volume of $4\cdot10^{-2}\,\text{mm}^3$, has been achieved using a different protocol based on an auxiliary frequency tone, without Dynamical Decoupling \cite{wangSCIADV2022}. Typical concentration sensitivity (\textit{i.e.} sensitivity normalized over the square root of the spin density \cite{balasubramanianNANOLETT2019,taylorNATPHYS2011}) values reported for NV centers are on the order of few units up to ten $n \text{T} /\sqrt{\text{Hz}}\,\mu\text{m}
^{3/2}$ \cite{balasubramanianNANOLETT2019,zhouPRX2020}.   
 
Molecular spins have shown sufficiently long coherence times to observe Rabi oscillations even at room temperature and with relatively high spin concentrations (typically from 0.01 up to few \%, corresponding to $\approx$ 100 up to 1 ppm)
\cite{schaefterADVMAT2023,yamabayashiJACS2018,atzoritesiJACS2016,atzorimorraJACS2016,baderPhysChemChemPhys2017,baderNATCOMM2014,yuJACS2016}. Quantum correlation -entanglement- between molecular spins has been observed in static configurations \cite{candiniPRL2010,garlattiNATCOMM2017,timcoNATTECH2009} while schemes to entangle spins by exploiting excited molecular levels have been proposed \cite{troianiPRL2005,luisPRL2011}. Due to these intrinsic quantum properties, molecular spins are attracting much interest for the implementation of quantum gates and algorithms \cite{ghirriMAGNETOCHEM2017,nakazawaAngChemIntern2012,chizziniPHYSCHEMCHEMPHYS2022,Chiesa2021,Chiesa2020,luisPRL2011,ranieriCHEMSCI2023,ranieriANGCHEMINTED2023}. The possibility to tailor the external organic ligands \cite{schaefterADVMAT2023,ferrandosoriaNATCOMM2016,nakazawaAngChemIntern2012}, and to selectively link/attach spin centers to a desired target, being this either an analyte in a supramolecular structure or a functional surface, is also well documented \cite{serranoNATCOMM2022,serranoNATMAT2020,candiniSciRep2016}. As matter of facts, this turns out to be one of the key characteristic of molecular spins \cite{yuACSCENTRSCI2021}. This provides a valid alternative to study biological systems as well as functional surfaces such as topological 2D materials or superconductors \cite{serranoNATCOMM2022,serranoNATMAT2020}.
Despite this peculiarity may actually open new avenues for quantum technologies \cite{troianiJMMM2019}, except for some interest and curiosity \cite{yuACSCENTRSCI2021,maniCHEMPHYSREV2022} and theoretical proposals \cite{mullinPHYSREVRES2023,schaffryPRA2010}, there are no experimental reports concerning the realization of quantum sensing schemes with magnetic molecules yet. 

In this work we consider two different classes of molecular spin centers, i) single transition metal ions, namely an oxovanadium tetraphenyl porphyrin (VO(TPP) for short), and ii) organic radicals, namely the $\alpha,\gamma$-bisdiphenylene-$\beta$-phenylally (BDPA for short), as two prototypical systems for testing the sensing of AC magnetic fields under Dynamical Decoupling-based magnetometry. The applied Radiofrequency (RF) fields (here with frequency values between $\approx$ 0.5 and 1 MHz and variable amplitude and phase) are synchronized to MW pulse sequences such Hahn echo and Dynamical Decoupling \cite{hirosePRA2012}. The sensing principle is based on the phase accumulation induced by the RF field during the free spin precession time of the protocols \cite{degenREVMODPHYS2017}. The variation of the phase of the echo signal measured at the end of the protocol can be related to the amplitude and the phase of the RF field applied, demonstrating the potential of molecular spin ensembles for quantum sensing for the first time. Moreover, the possibility to perform sensing of magnetic fields and the unique peculiarities of molecular spins, paves the way for the implementation of noise spectroscopy \cite{zapasskiiADVOPTPHOT2013,aleksandrovJPHYSCONF2011,zhaoPRA2014} or for using spins as local probes, once integrated into hosts, such as metallic or superconducting surfaces.

\section{Sensing Principle}

\begin{figure}[h]
\centering
\includegraphics[width=1\textwidth]{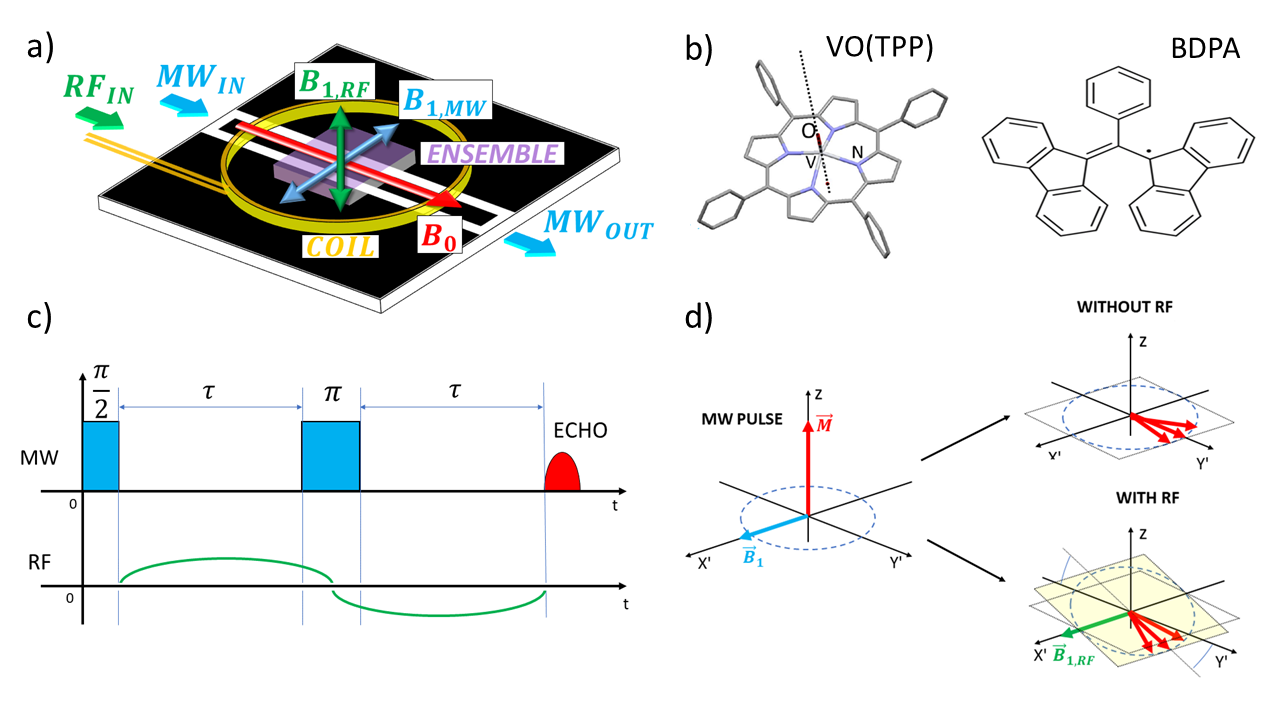} 
\caption{a) Sketch of the coplanar resonator used in the experiment with the position of the sample (purple) and the radiofrequency (RF) coil (orange) used. Red, green and blue arrows show the orientations of the static magnetic field ($\textbf{B}_{0}$), the microwave (MW) field ($\textbf{B}_{1,MW}$), and the RF field ($\textbf{B}_{1,RF}$), respectively. b) X-ray structure of the VO(TPP) molecule and sketch of the BDPA molecule used in this work. c) Timing diagram of the protocol used for quantum sensing. The MW sequence is the Hahn echo and the RF modulation is synchronized with its free precession time, $\tau$. d) Sketch of the precession of the spins in the rotating reference frame during the application of a MW pulse and, then, with or without the application of the RF field (see labels). The different precession plane when the RF modulation is added is shown in yellow.}
\label{Figure1}
\end{figure}

Sensing of DC or AC fields requires different protocols. DC magnetometry with spins is based on the slightly change of the resonant condition given by the addition of the field to be probed, which (vector) sums to the static one. Typical experiments are performed by measuring the change in the oscillations of the Ramsey fringes or the Rabi oscillations induced by the presence of the additional DC field \cite{degenREVMODPHYS2017}. However, these methods can be hardly used for probing slow-varying fields or the local environment surrounding the spins, since they would give only a time-averaged (eventually zero) response. Conversely, AC magnetometry \cite{degenREVMODPHYS2017}, which is considered in our work, can be applied to probe oscillating magnetic fields in a relatively large frequency band, and it can be further extended also to the multifrequency case. 

The full prototypical Hamiltonian of a quantum sensing experiment is

\begin{equation} 
H=H_{0}+H_{control}+H_{V},
\label{eq_full_Ham}
\end{equation}  

where $H_{0}$ is the Hamiltonian describing the system acting as a sensor, $H_{control}$ is the protocol used on the sensor to drive it and measure its response, while $H_{V}$ is the Hamiltonian describing the physical quantity to be probed.
In our experimental case, $H_{0}$ is the Hamiltonian describing the VO(TPP) \cite{bonizzoniNPJQUANT2020} or the BDPA \cite{bonizzoniAPPLMAGNRES2023} spin ensemble, while $H_{control}$ is the microwave pulse sequence (Hahn echo or Dynamical Decoupling) used to induce the spin free precession and to get a spin echo. $H_{V}$ describes the interaction induced by the RF oscillating magnetic field $\textbf{B}_{1,RF}$ acting during the free spin precession, which is the signal to be measured. 
In our scheme (see Fig. \ref{Figure1}), the microwave magnetic field $\textbf{B}_{1,MW}$ lies on a plane perpendicular to the direction of the static one $\textbf{B}_{0}$, allowing for the perpendicular excitation of the spin transitions \cite{abrambleaney}. The direction of the RF field is perpendicular to the static one, as for typical magnetic field sensing protocols \cite{zhouPRX2020,wrachtrupJMR2016,taylorNATPHYS2011,balasubramanianNATMAT2009}.
On resonance and in absence of RF excitation, the effect of the microwave sequence is to induce the free spin precession, which can be visualized as a spin dephasing in the xy plane of the rotating reference frame, as in Fig. 1.d \cite{abrambleaney}. The RF modulation acts only when the microwave field is off and (vector) sums to the static magnetic field modulating the resonance condition and the direction around which spin precession occurs. In the rotating reference frame this modulates the orientation of the precession plane with respect to the readout one, which remains the same as in absence of RF modulation, leading to a change in both amplitude and phase of the echo, as in Fig. \ref{Figure1}.d. If the RF frequency is synchronized with the MW sequence, these effects build up during the free spin precession, leading to an extra phase accumulation in the preceeding spins and to a macroscopic change (reduction) of the echo amplitude, which can be used to estimate the RF field \cite{taylorNATPHYS2011,balasubramanianNATMAT2009}.\\   

To better understand the phase accumulation given by the RF modulation we focus on the simplest protocol used in this work, \textit{i.e.} the Hahn echo sequence, which is implemented by giving a first $\pi/2$ pulse followed by an interpulse delay $\tau$  and, then, by a $\pi$ pulse. An electron spin echo is found after a delay $\tau$ from the $\pi$ pulse, as in Fig. \ref{Figure1}.c. The RF excitation is a monochromatic field in the form $B_{1,RF}$(t)= $B_{1,RF}$ $sin(2\pi \nu_{RF} t+ \phi_{RF})$ with amplitude $B_{1,RF}$, frequency $\nu_{RF}$ and additional phase term $\phi_{RF}$. The RF modulation is synchronized with the period of the free precession time by choosing $\nu_{RF} = n /(2 \tau)$ (with $n$ as an integer, non negative, number). Under the conditions in which the RF field is small with respect to the static one, the RF field modulates the Zeeman energy of spins and can be taken as $H_{V}= \textit{g}\,\mu_{B} B_{RF}(t)$, which gives a corresponding angular frequency $\omega_{V}= \frac{H_{V}}{\hbar} = \frac{ \textit{g}\,\mu_{B} B_{RF}(t)}{\hbar}$. The total phase accumulation after a single Hahn echo sequence is \cite{taylorNATPHYS2011,balasubramanianNATMAT2009}:

\begin{equation} 
\begin{split}
\phi_{RF} = \int^{2 \tau}_{0} \omega_{V} dt = \int^{2 \tau}_{0} \frac{\textit{g} \mu_{B}}{\hbar} B_{RF}(t) dt = \int^{2 \tau}_{0} \frac{\textit{g} \mu_{B}}{\hbar} B_{1,RF}sin(2\pi \nu_{RF} t + \phi_{RF}) dt  = \\ = 2 \int^{\tau}_{0} \frac{\textit{g} \mu_{B}}{\hbar} B_{1,RF} sin(2\pi \nu_{RF} t +\phi_{RF}) dt =  \frac{2\textit{g} \mu_{B}}{\hbar} B_{1,RF} \int^{\tau}_{0}  sin(2\pi \nu_{RF} t +\phi_{RF}) dt   \propto B_{1,RF}
\end{split}
\label{eq_phase_acc}
\end{equation}  
 
The phase accumulated by the echo signal is proportional to the amplitude of the applied magnetic field $B_{1,RF}$ and it also depends by the integral of the RF modulation over the free precession time of spins, \textit{i.e.} on the synchronization (phase and duration) of $\textbf{B}_{1,RF}$ with respect to the sequence. Next, we design our experiments in order to show the dependences of the echo phase on the RF amplitude, RF phase and on the integral of the modulation.

\section{Experimental Results}
\label{sec_results_hahn}

We test the protocol shown in Fig. \ref{Figure1}.c in several experiments performed on the VO(TPP) sample.


\begin{figure}[h!]
\centering
\includegraphics[width=1\textwidth]{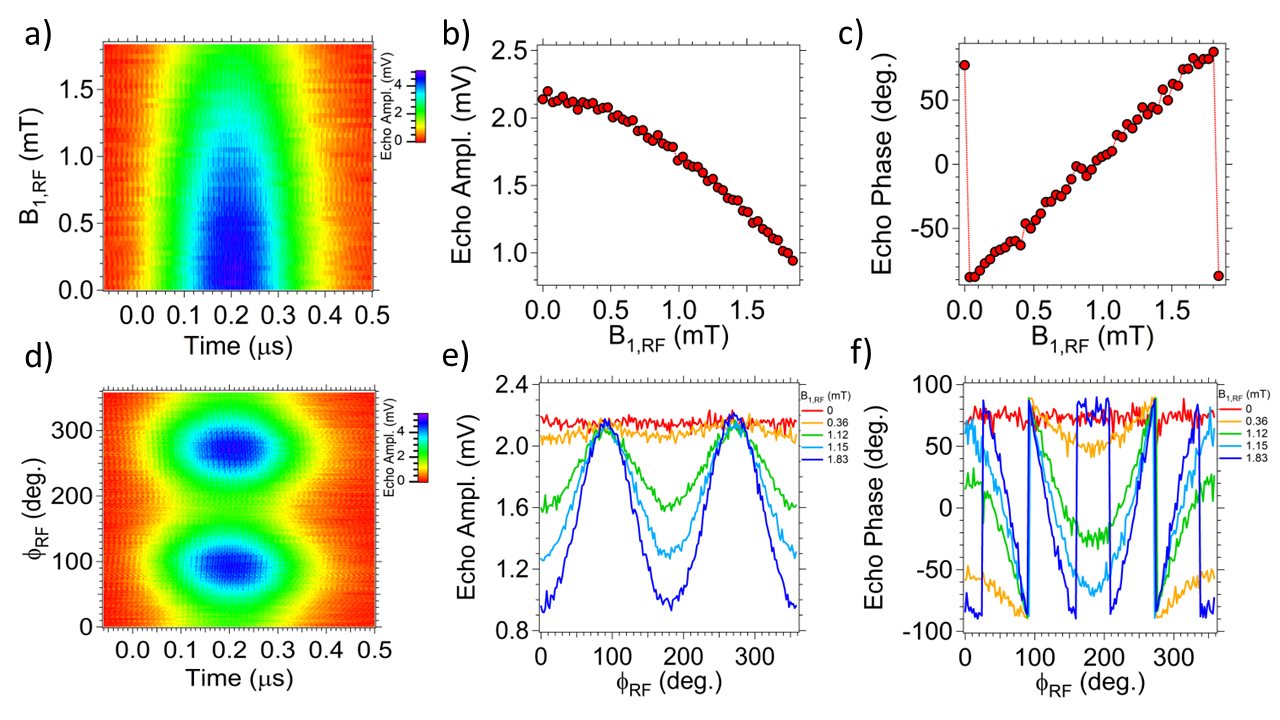}
\caption{a,b,c) Dependence of Hahn echo on the amplitude of the RF magnetic field for the VO(TPP) sample. a) 2D map of the echo signal measured in time domain for different amplitudes of RF magnetic field. Echo amplitude (b) and phase (c) extracted from the map in (a) as a function of the RF amplitude. The phase of the output echo, which is wrapped between -90 and 90 degrees for better clarity, linearly increases  with the RF amplitude. d,e,f) Dependence of Hahn echo on the phase of the RF magnetic field for the VO(TPP) sample. d) 2D map of the echo signal measured in time domain for different phase values of RF magnetic field. Echo amplitude (e) and phase (f) extracted from the map in (d) as a function of the RF phase.}
\label{Figure2}
\end{figure}

We first fix $n=1$ and $\phi_{RF}=0$ in the RF modulation (see Eq. \ref{eq_phase_acc}). The RF amplitude $B_{1,RF}$ is increased step-by-step and the Hahn echo signal is acquired for each field amplitude value. The Hahn echo amplitude and phase extracted as a function of the amplitude of the RF magnetic field are shown in Fig. \ref{Figure2}.a,b,c. The echo amplitude decreases as the RF amplitude is increased, while the echo phase linearly increases with the RF amplitude. Incidentally, this latter result is quite similar to what observed on Nitrogen-Vacancy centers by using similar protocols  \cite{taylorNATPHYS2011,hirosePRA2012,woodPRB2018}, and it nicely shows the effect of the phase accumulation induced by the RF field on the spin precession, according to Eq. \ref{eq_phase_acc}. Here the echo amplitude never vanishes and the phase exhibits a linear dependence on the amplitude of the RF field. Similar results are observed also if the interpulse delay is changed between 900 ns and 1700 ns (See Supplementary Information).  


Secondly, we investigate the effect of the phase of the RF field, $\phi_{RF}$. In the Bloch sphere (rotating reference frame) this operation corresponds to set a different orientation of the RF magnetic field in the $xy$ plane with respect to the direction of the microwave pulses (which are along the $+ x$ axis, since $\phi_{\pi/2}=\phi_{\pi}=0$ \cite{bonizzoniPRAAPPL2022}). This introduces an initial phase delay between the Hahn echo sequence and the RF modulation itself, changing the position at which the maximum of the magnetic field occurs during the free precession time. Results for a full 0 to 360$^{\circ}$ phase sweep for different $B_{1,RF}$ values are shown in Fig. \ref{Figure2}.d,e,f. A modulation of the echo amplitude with period of $180^{\circ}$ is visible as a function of the RF phase (Fig. \ref{Figure2}.d. e). The effect of the applied modulation is maximum for $\phi_{RF}=0, 180^{\circ}\,\text{\,and\,}360^{\circ}$, while is completely negligible for $\phi_{RF}=90^{\circ}\text{ and }270^{\circ}$, where the echo amplitude is equal to the $B_{1,RF}=0$ case. The echo results to be sensitive to the orientation of the RF field in the precession plane. 
This effect is more evident in the echo phase, which undergoes a larger variation with respect to the $B_{1,RF}=0$ case as $B_{1,RF}$ is increased. The slope of the phase signal as a function of $\phi_{RF}$ increases with $B_{1,RF}$ showing that faster phase oscillations between $-90^{\circ}$ and $+90^{\circ}$ are achieved for larger RF field amplitudes. Similar oscillation are visible also if the interpulse delay is changed between 800 ns and 1900 ns (see Supplementary Information).
We explain these results according to Eq. \ref{eq_phase_acc}. From a mathematical point of view, changing the RF phase is equivalent to change of the extremes of the integration interval by an amount $t=-\phi_{RF}/(2\pi \nu_{RF})$. This implies that the integral goes from a maximum value $\phi_{RF}=0$ down a minimum $\phi_{RF}=90^{\circ}$ (RF modulation antisymmetric over $\tau$, integral equal to zero, no effective phase accumulation) and, then, up to another maximum for $\phi_{RF}=180^{\circ}$. These results show that the echo phase signal is sensitive to the integral of the RF modulation applied during the free precession time.\\

\begin{figure}[h!]
\centering
\includegraphics[width=1\textwidth]{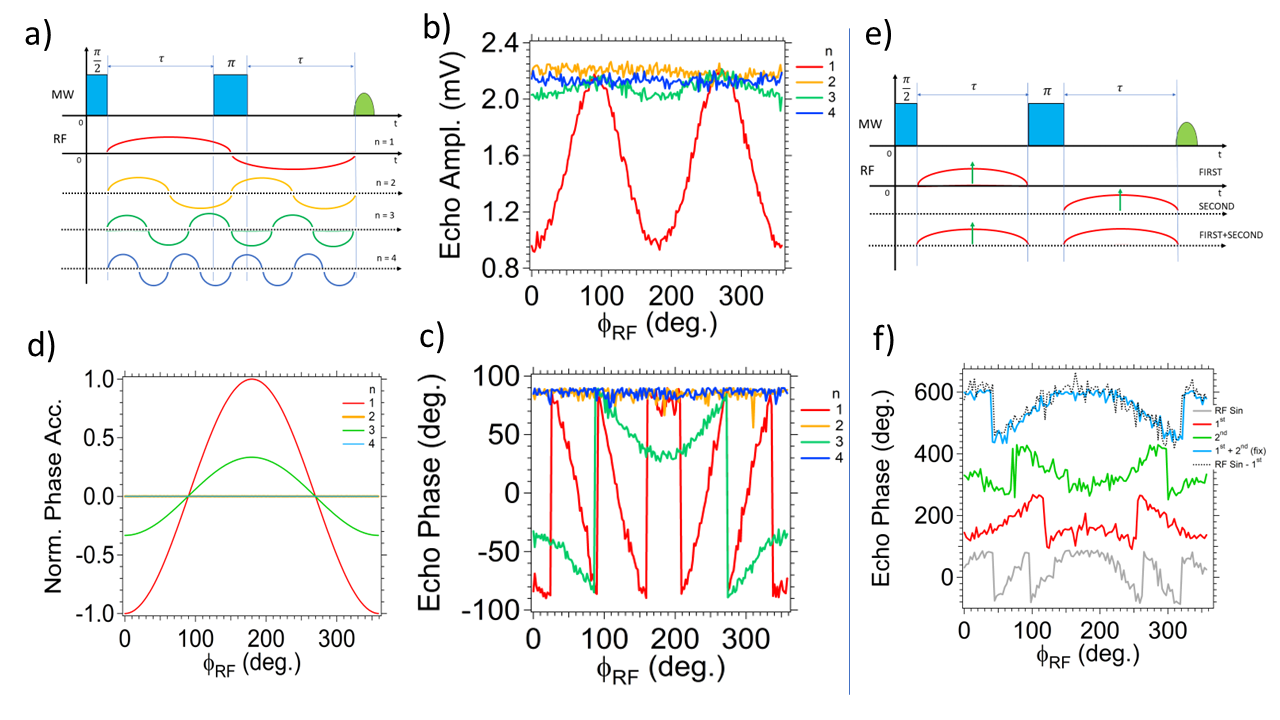}
\caption{a) to d) Investigation of the effect of the symmetry of the RF field for VO(TPP) sample. a) Sketch of the protocol used, echo amplitude (b) and phase (c) as a function of the RF phase $\phi_{RF}$. d) simulation of the phase accumulation carried out by means of Eq. \ref{eq_phase_acc}. e) and f) Investigation of the sum of RF modulations for VO(TPP) sample. e) sketch of the protocol used. f) echo phase measured as a function of the RF phase, $\phi_{RF}$.}
\label{Figure3}
\end{figure}


The next set of experiments further assess the dependence of the echo signal on the integral of the phase accumulation, according to Eq. \ref{eq_phase_acc}. 
We first change the symmetry of the RF modulation during $\tau$, \textit{i.e.} the factor $n$ of $\nu_{RF}$ for fixed $B_{1,RF}=1.8\,mT$. In this way, the modulation signal becomes symmetric (even $n$) or antisymmetric (odd $n$) over the duration $\tau$, as in Fig. \ref{Figure3}.a. Results as a function of $\phi_{RF}$ are shown in Fig. \ref{Figure3}.b,c for VO(TPP) sample (here we have $\tau = 1200\,ns$, corresponding to $\nu_{RF}=\,n\,0.83\,$MHz). The echo amplitude shows a modulation for $n=1,3$ (0.83 and 2.5 MHz, respectively), similar to the one in Fig. \ref{Figure2}.e, while a negligible modulation is found for $n=2,4$ (1.7 and 3.3 MHz, respectively). The amplitude modulation for $n=3$ is smaller with respect to the one for $n=1$, suggesting that the RF signal has lower effect. The result is more evident in the echo phase signal, where periodical oscillations are clearly visible only for odd $n$ and the signal is unperturbed for even $n$. In addition, the echo phase for $n=1$ shows faster oscillations as a function of $\phi_{RF}$ with respect to $n=3$, corroborating the different effect on the same total precession time.
These results are further corroborated by the simulation performed according to Eq. \ref{eq_phase_acc} (Fig. \ref{Figure3}.d). In particular the phase accumulation predicted is zero for odd $n$ only for $\phi_{RF}=90^{\circ},270^{\circ}$, while is always zero for all even $n$. Also the relative scaling between $n=1$ and $n=3$ is predicted, highlighting the different effectiveness of the two modulations.


Then, we further show how the phase accumulation builds up during the free precession time $2\tau$ by applying half period of RF modulation only during the first $\tau$ interval, only during the second one or, alternatively, the same half period of modulation in both $\tau$ intervals, as shown in Fig. \ref{Figure3}.e. Sweeping the RF phase only during the first or the second $\tau$ interval (red trace, \emph{first}, and green trace, \emph{second}, respectively) at fixed $B_{RF}=1.8\,$mT gives phase oscillation as in Fig. \ref{Figure2} but with slower changes ($\approx$ doubled period) with respect to the full RF modulation applied (grey trace, \emph{RF sin} in the legend). The nearly opposite sign further confirms the different orientation of the modulation in the \textit{xy} precession plane. Finally, if the RF modulation is applied on both $\tau$ intervals ($B_{RF}=1.8\,$mT) and the phase of the first one is swept (for fixed $\phi_{RF}=0\,^{\circ}$ on the second one) the behaviour of the echo phase results from the difference of the two modulation integrals over $\tau$. This can be realized by plotting the difference between the full RF modulation (grey trace) and the first one (red trace), which gives the contribution which is "removed" from the total RF modulation due to the effect of the first one (black dashed trace, to be compared with light-blue one). These results corroborate the algebraic additivity of the modulations over the two different $\tau$, as expected from Eq. \ref{eq_phase_acc}. 
Results similar to the ones reported in Figs. \ref{Figure2},\ref{Figure3} hold also for the BDPA sample (see Supplementary Information), suggesting that the protocols tested can be successfully extended to other molecular spins classes.\\


\begin{figure}[h!]
\centering
\includegraphics[width=1\textwidth]{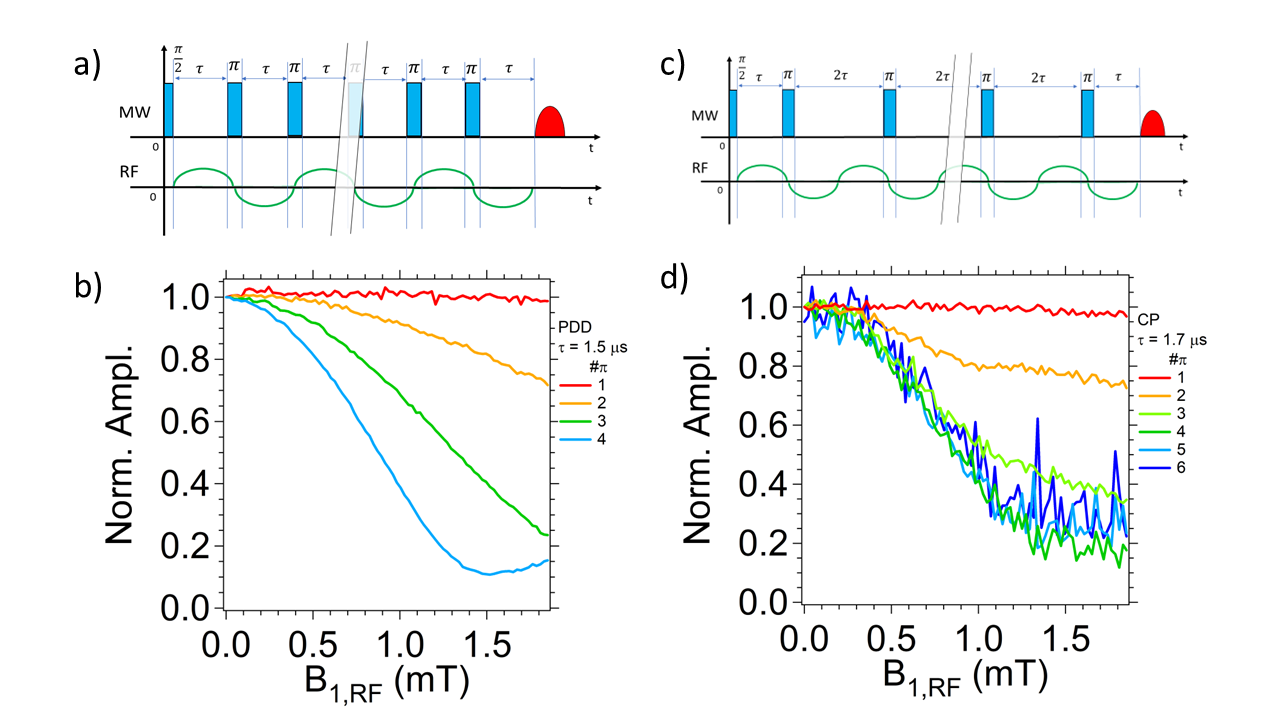}
\caption{a) and c) Sketch of the Dynamical Decoupling protocols used in this work: the Periodic Dynamical Decoupling (PDD,a) and the Carr-Purcell-Meiboom-Gill sequence (CP). Normalized echo amplitude as a function of the RF field amplitude for different number of $\pi$ pulses used in the protocols. Periodic Dynamical Decoupling (c) and Carr-Purcell-Meiboom-Gill sequences (d) are shown, respectively.}
\label{Figure4}
\end{figure}

Dynamical Decoupling protocols \cite{souzaPRL2011} can be also exploited for quantum sensing \cite{degenREVMODPHYS2017,hirosePRA2012,souzaPHYLOSRCS2012}. These, essentially, increase the spin lifetime and also gives longer time for the RF field to act on spins during free precession, enhancing the sensitivity. To this end, we extend the  protocol of Fig. \ref{Figure1} to Dynamical Decoupling sequences, and we apply it on the BDPA organic radical, since its memory time is found to largely increase through a Carr-Purcell-Meiboom-Gill sequence performed through the coplanar resonator (see Supplementary Information). Here, we notice that applying a Dynamical Decoupling protocol implicitly acts as bandpass frequency filter to the environmental field (noise) acting on spins \cite{degenREVMODPHYS2017,hirosePRA2012,souzaPHYLOSRCS2012}. This implies that, in order to be effective, the applied RF modulation must match the band of the filter (\textit{i.e.} being synchronized with the protocol and not suppressed by the response of the filter itself). In other words, the sensitivity strongly depends upon the specific choice of parameters used in the Dynamical Decoupling (frequency $\nu = 1/ \tau$, phase of the RF modulation and number of $\pi$ pulses) \cite{taylorNATPHYS2011,phamPRB2012}.


We first investigate a Periodic Dynamical Decoupling (PDD) \cite{hirosePRA2012}. Here, a first $\pi/2$ pulse is followed by a train of $\pi$ pulses equally-spaced by in time a delay $\tau$. The RF field has period $T=1/(2\tau)$ and $\phi_{RF}=0$. In this way the RF modulation builds always up after the spin refocusing given by each $\pi$ pulse. The echo considered and analyzed in this protocol is the one appearing at a time $\tau$ after the last $\pi$ pulse sent. Results are shown in Fig. \ref{Figure4}.b. The echo amplitude normalized over each corresponding zero modulation value decreases as the number of $\pi$ pulses increases, due to the longer time available to the modulation to act on spins. 


Next, we consider a Carr-Purcell-Meiboom-Gill protocol (hereafter, CP) \cite{degenREVMODPHYS2017,hirosePRA2012}. Here a first $\pi/2$ pulse is followed by an initial interpulse delay $\tau$ and, then, by a train of $\pi$ pulses equally-spaced by $2\tau$. Again, we considered the echo appearing at a time $\tau$ after the last $\pi$ pulse sent, and the RF modulation has a period $T=1/(2\tau)$ and $\phi_{RF}=0$ in order to synchronize with every spin refocusing. We have investigated this dependency for different interpulse delays ($\tau = 1, 1.3, 1.7\,\mu$s, corresponding to $\nu_{RF}=1,0.77,0.59\,$ MHz respectively) and number of $\pi$ pulses, and we report in Fig. \ref{Figure4}.d the results obtained for $\tau=1,7\,\mu$s. Additional results are reported in Supplementary Information.
The echo amplitude normalized over each corresponding zero modulation value decreases as the number of $\pi$ pulses is increased. However, for larger number of $\pi$ pulses applied the echo signal gets smaller, and the addition of the RF modulation results in a strong decrease of its signal and of the \emph{signal-to-noise} ratio, preventing from an accurate measure of both echo amplitude and phase under our experimental conditions. We attribute this effect to the increasing inhomogeneity of the magnetic field introduced by the RF coil, which induces strong dephasing during spin precession.

\section{Discussion}
 
Our results demonstrate that the echo signal given by the spins can be used to probe an oscillating monochromatic magnetic field applied perpendicularly to the static one. 

We first discuss the magnetic field sensitivity achieved under our experimental conditions. 
To this end, we consider the linear dependence of the echo phase shown in Fig.\ref{Figure2}.c. This data allow us to estimate a transduction coefficient, $\frac{d \phi_{echo}}{d B_{RF}}$, between the echo phase and the RF magnetic field amplitude through

\begin{equation}
\frac{d \phi_{echo}}{d V_{RF}} = \frac{d \phi_{echo}}{d B_{RF}} \frac{d B_{RF}}{d V_{RF}}, 
\label{eq_phase_deriv}
\end{equation}

where the relation between the amplitudes of RF magnetic field and RF voltage given by the AWG, $\frac{d B_{RF}}{d V_{RF}}$, is known from independent calibrations (see Supplementary Information). Eq.(\ref{eq_phase_deriv})  refers to the "slope detection" case defined in \cite{degenREVMODPHYS2017}.  From the data of Fig. \ref{Figure2}.c we can get $\frac{d B_{RF}}{d \phi_{echo} }=(\frac{d \phi_{echo}}{d B_{RF}})^{-1}=9.8\cdot 10^{-6}\,T/^{\circ}$ for VO(TPP) when the Hahn echo sequence is used to probe an RF magnetic field with frequency $\nu_{RF}=1/(2\tau)=0.42\,$MHz.  Under our experimental conditions and with our experimental set up, the  phase resolution (\textit{i.e.} the minimum phase variation of the echo signal which can be detected) is $\approx\,1^{\circ}$ \cite{bonizzoniPRAAPPL2022}, which implies a minimum detectable field amplitude of $B_{min}=9.8\cdot 10^{-6}\,T/^{\circ}\,1^{\circ}=9.8\cdot 10^{-6}\,T$. To get a direct comparison of the sensitivity achieved for similar experiments performed on spin ensembles, we evaluate the sensitivity as the minimum detectable field for unitary \emph{Signal-to-noise} ratio and unitary acquisition bandwidth \cite{blankJMR2017}. This reads as $S \approx 6\cdot 10^{-6}\,T/\sqrt{\text{Hz}}$ (see Supplementary Information), which well compares with some reports in the literature \cite{yahataAPL2019,woodPRB2018}.
In our experiments the effective number of spins is of the order of $\approx\,10^{14}$ spins, while the number of spins which are taking part to the Hahn echo signal is $\approx\, 10^{12}-10^{13}$ (see Supplementary Information). This gives the size (in absolute number of spins) of the sensor which is coherently experiencing the RF modulation applied. Taking the values of the spin concentration ($\rho=2.3\cdot10^{19}\,spin/cm^3$ \cite{yamabayashiJACS2018,drewINORGCHIMACTA1984}) we can estimate an active sensing volume $V_{s}=1.75\cdot 10^{-3}\, \text{mm}^3$ for VO(TPP) and obtain a concentration sensitivity $S_{vol}=S/\sqrt{\rho} \approx 1.2\cdot 10^{-9}\,T\,\mu m^{3/2}/\sqrt{\text{Hz}}$. This value is consistent with the theoretical limit of sensitivity expected from the derivation in \cite{degenREVMODPHYS2017} (see Supplementary Information), for which the lower bound to sensitivity results to be $S_{min}\approx 1.5\cdot 10^{-6} T/\sqrt{\text{Hz}}$, corresponding to an average concentration sensitivity $S_{min,vol}\approx 3.1\cdot 10^{-10} T \mu m^{3/2}/\sqrt{\text{Hz}}$. 

\begin{figure}[h]
\centering
\includegraphics[width=1\textwidth]{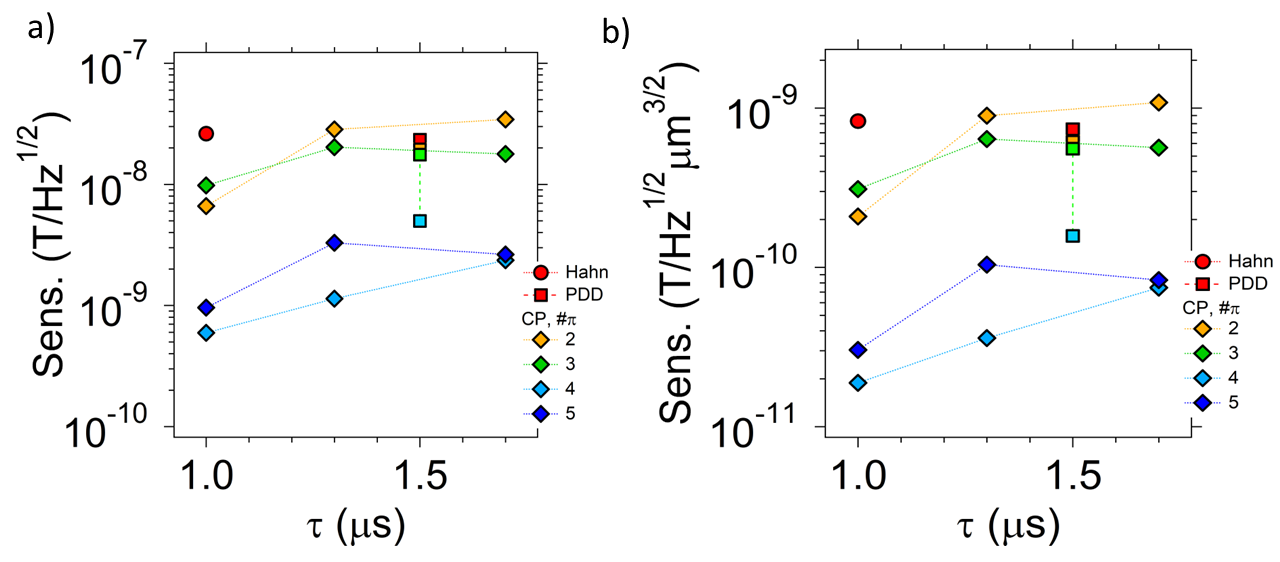}
\caption{Best Sensitivity (a) estimated from the sensing experiments with Dynamical Decoupling protocols on BDPA sample and its corresponding Sensitivity per unit of concentration (b). For PDD the colorscale indicates the different number of $\pi$ pulses used from 1 (red) up to 4 (cyan). Point for Hahn echo (red, corresponding to CP1 and PDD1) is added for comparison.}
\label{Figure5}
\end{figure}

Likewise, we estimate the  sensitivity for the Dynamical Decoupling experiments by using the same definition as above. To take into account the different behaviour of BDPA as a function of $B_{1,RF}$, we use the maximum of the first derivative of the phase as a function of $B_{1,RF}$ in the estimation of the minimum magnetic field. This ensures that the definition is equivalent to the one used for VO(TPP) if the phase depends linearly by the RF magnetic field, while the maximum signal variation is considered if the data are in the non-linear regime (see Supplementary Information). Our results are summarized in Fig. \ref{Figure5}.

Overall, for a given $\tau$, increasing the number of $\pi$ pulses (vertical alignment of points) increases the sensitivity to the RF magnetic field in all cases. PDD protocol results to have similar sensitivity with respect to CP sequences for 2 and 3 $\pi$ pulses, while becomes less sensitive for larger number of pulses. This matches well with the different total precession time available for sensing the RF field. Our estimated sensitivity values ($10^{-9} T/\sqrt{\text{Hz}}$) compare well with the values reported in the literature for ensembles of NV centers in diamond \cite{phamPRB2012,rondinREPPROGRPHYS2014,taylorNATPHYS2011,farfurnikJOURNOFOPT2018,zhouPRX2020,balasubramanianNANOLETT2019} and, for instance, they outperform what recently reported for defects in hexagonal Boron Nitride (hBN) layers \cite{rizzatoNATCOMM2023} or spin-vacancy nanodiamonds \cite{tsukamotoSCIREP2022,aslamNATREVPHYS2023}. It is worth to notice that, in spite of the fact that optical detection can be pushed to the single event (\textit{i.e.} single photon), the detection of photons typically has low yield. This in part compensates the advantage given by the longer memory time of NV center in diamond as compared to that of molecular spins. However it is also remarkable that the sensitivity here reported for molecular spins can be reached with a relatively low number (4-5) of applied $\pi$ pulses, without optical detection \cite{mullinPHYSREVRES2023}, and by using samples with higher spin concentration with respect to the typical values used in literature for defects in solids. This suggests that there are still margins of improvement for the sensitivity obtained with molecular spins.  

To get more insights on our results, we first note that the performances obtained in this type of experiments depend not only on the sensor itself, but also by the whole detection chain, including the microwave resonator and microwave electronics used to read out the echo signal, which can be potentially improved in future experiments. On the other hand, the RF field  $\textbf{B}_{1,RF}$ is applied over the whole volume in such a way that each spin is locally experiencing the same RF field. This implies that the transduction coefficient estimated with Eq.(\ref{eq_phase_deriv}) is the same for all spins, and it can be expected to hold also for single, isolated spins using the phase accumulation method and the same experimental protocol used in this work. To compare $\textbf{B}_{1,RF}$ in our experiments with local fields, let us consider the magnetic field generated by dipole-like spin center with a magnetic moment equal to Bohr's magneton (\textit{e.g.} a electronic spin S=1/2). The field at distance of 5 nm from the axis is $B_{S}=7 \cdot\,10^{-6}\,$T which is comparable to what we used in our experiments. This suggests that a local magnetic field or magnetic moment would be detectable in an experiment in which a single VO(TPP) molecular sensors is used at nanoscale distance \cite{mullinPHYSREVRES2023}. 
This rough estimation suggests us a strategy for realistic sensing experiments at the molecular scale, in which single molecular sensors are attached and interact with magnetic analytes or magnetic surfaces or to functional magnetic surfaces through a selective organic thread. For instance, we expect that electromagnetic noise locally generated by molecular process can be detectable by an electron spin such as a radical or a VO(TPP) spin center. Experiments can be performed on diluted ensembles of units containing the analyte bringing spin label and repeating the quantum sensing protocol (e.g. DD) at different frequencies in such a way that noise spectrum can by reconstructed by Fourier analysis. Such noise spectroscopy has been developed in the context of magnetic resonance \cite{slichterbook,zapasskiiADVOPTPHOT2013,aleksandrovJPHYSCONF2011} and also proposed for NV centers \cite{zhaoPRA2014}.
Finally, another potential benefit of the introduction of a local molecular spin sensors relies on the possibility to probe the magnetic properties of Electron Spin Resonance-silent ions or of moieties with large zero field splitting, which cannot be directly or easily studied by Spin Resonance techniques \cite{komijaniPHYSREVMAT2018}.

\subsection{Conclusions}

In conclusion, we have realized quantum sensing protocols using ensembles of molecular spins as sensing element and embedded into a planar microwave resonator. To assess the performances of our sensor, we implement the detection of RF magnetic fields within continuous Dynamical Decoupling magnetometry schemes.
The sensitivity achieved with Hahn echo sequence is $S\approx 10^{-8} -10^{-6}\,\text{T}/\sqrt{\text{Hz}}$, corresponding to concentration sensitivity $S_{vol}\approx 10^{-10}\,\text{T}\,\mu \text{m}^{3/2}/\sqrt{\text{Hz}}$. Extending our experiments with Dynamical Decoupling protocols, such as the Period Dynamical Decoupling and Carr-Purcell-Meigboom-Gill, we can improve the sensitivity  up to $S \approx 10^{-10}-10^{-9}\,\text{T}/\sqrt{\text{Hz}}$ ($S_{vol}\approx 10^{-11}\,\text{T}\mu \text{m}^{3/2}/\sqrt{\text{Hz}}$) with a relatively low (4-5) number of $\pi$ pulses. This sensitivity compares well with the standard ones reported for NV centers and our concentration sensitivity results to be even slightly better due to the different spin density of our sensors. Our scheme is based on the resonant readout of a spin echo and does not require optical (fluorescence) readout, which usually has the drawback of having low photon conversion efficiency. Moreover, similar sensitivity values are achieved with rather short sequences (few pulses, to be compared with tens or hundreds) making the sensing protocol easier to be implemented.  
We have shown that molecular spins can be competitive solid-state spins for performing the sensing of AC magnetic fields (with frequencies up to 1-2 MHz in our experiments). Our results can find application in sensing the amplitude or the phase of periodic signals (here, RF magnetic fields) or in noise spectroscopy to detect local processes using molecular (electronic) spin labels. 
Our results can be extended to oscillating fields containing multiple Fourier components thus to B$_{1,RF} (t)$ with arbitrary time dependence \cite{degenREVMODPHYS2017,hirosePRA2012}.
We finally notice that the phase additivity over the free spin precession time might be useful in designing protocols for the Revival of Silent Echo (ROSE) of spin ensembles \cite{ranjanPRL2022,damonNEWJOP2011} or for the realization quadrature detection \cite{levensonSuperCondSciTech2016}.

\begin{acknowledgments}
We thank Dr. Johan van Tol (National High Magnetic Field Laboratory, Florida, USA) for the preparation of BDPA samples.\\
This work was funded by the H2020-FETOPEN ”Supergalax” project (grant agreement n. 863313) supported by the European Community.\\
\end{acknowledgments}

\paragraph{Author Contributions}
CB, AG, MA conceived the experiments. CB prepared the resonator, the experimental set up and carried out all measurements and data analysis. FS prepared the VO(TPP) sample. The manuscript was written by CB with inputs from all authors. CB, AG, MA contributed to the discussion of the results. The manuscript has been revised by all authors before submission.

\paragraph{Competing Interests}
All authors declare no competing interests.

\section{Methods}

\subsection{Samples and resonator}
We use two different molecular spins centers (Fig. \ref{Figure1}.b). The first one is a 2\% doped crystalline sample of VO(TPP) in its isostructural diamagnetic analog, TiO(TPP). Each molecule has an electronic spin $S\,=\,1/2$ ground state and an additional hyperfine splitting given by the interaction with the $I=7/2$ nuclear spin of the $^{51}$V ion (natural abundance: 99.75\%). The magnetic properties and the electron spin resonance spectroscopy of this molecule have been previously reported in \cite{yamabayashiJACS2018}. The latter one is a BDPA organic radical diluted into a Polystirene matrix with spin concentration of $\approx,1\cdot10^{15}\,$, similar to the one reported in \cite{bonizzoniAPPLMAGNRES2023}. Each molecule has an unpaired electron giving an electronic spin $S=1/2$ center.
We place the sample on a superconducting coplanar resonator ($\nu_{0}\approx\,6.91$ GHz) made out of superconducting YBa$_2$Cu$_3$O$_7$ (YBCO) films on a Sapphire substrate (Fig. \ref{Figure1}.a), on which we have already integrated molecular spin ensembles \cite{bonizzoniSCIREP2017,ghirriAPL2015} and implemented microwave sequences for spin manipulation \cite{bonizzoniNPJQUANT2020,bonizzoniAPPLMAGNRES2023}. The Continuous Wave (CW) and Pulsed Wave (PW) microwave spectroscopy of VO(TPP) through the resonator has been previously reported in \cite{bonizzoniNPJQUANT2020,bonizzoniPRAAPPL2022}, while the one for BDPA has been previously reported in \cite{bonizzoniAPPLMAGNRES2023}. The sample and the resonator are cooled-down to 3 K into a commercial Quantum Design Physical Properties Measurement System (QD PPMS), which is also used to apply the external static magnetic field \cite{bonizzoniNPJQUANT2020,bonizzoniAdvPhys2018,bonizzoniAPPLMAGNRES2023}. A RF copper coil (diameter $\,=\,5\,mm$, height $\,=\,5\,mm$ and 7 windings) is placed on the surface of the resonator as in Fig. \ref{Figure1}.a, the way that the magnetic sample is at its central axis.
 
\subsection{Set up for the generation of sensing protocols}
The experimental setup is essentially the microwave heterodyne spectrometer previously reported in \cite{bonizzoniPRAAPPL2022,bonizzoniNPJQUANT2020} in which one channel of the waveform generator is used to generate the microwave excitation tone through frequency upconversion. The other channel of the waveform generator is used to generate a RF excitation which is routed directly to the copper coil with a dedicated RF coaxial line. In this way the setup can generate synchronized MW and RF pulse sequences in which each pulse parameter (amplitude, duration, phase, inter-pulse delay) can be tailored. The readout is performed at MW frequency and in time domain with an oscilloscope, after the output MW signal has been downconverted with a second mixer. The generation of the two excitations and acquisition are controlled by a home-written Python script.
Typical pulse parameters used in our experiments are: pulse durations between $t_{\pi/2}=80\,$ ns and $t_{\pi/2}=150\,$ ns for the $\pi/2$ pulse and between $t_{\pi}=160\,$ ns and $t_{\pi}=300\,$ ns for the $\pi$ pulse, and interpulse dely $\tau = 900\,$ ns up to $\tau = 1700\,$ ns. Each sequence is followed by a relaxation time $t_{relax}=15-20\,$ms to avoid sample saturation. The RF excitation amplitude can be as high as $V_{RF}=2.5 V$, limited by maximum output voltage available at AWG port, and a preliminary independent calibration procedure has been performed to convert $V_{RF}$ into its corresponding magnetic field $B_{1,RF}$ (see Supplementary Information).\\

\paragraph{Data Availability}
Experimental data are available from corresponding author upon reasonable request.


%

\end{document}